\documentclass[aps,prl,twocolumn,showpacs]{revtex4}

\usepackage{graphicx}%
\usepackage{dcolumn}
\usepackage{amsmath}

\makeatletter
\def\btt#1{\texttt{\@backslashchar#1}}%
\DeclareRobustCommand\bblash{\btt{\@backslashchar}}%
\makeatother

\begin{document}

\title[Short Title]{Nexus between  quantum criticality and the chemical potential \\pinning
in  high-$T_c$ cuprates}
 
 \date{\today}

\begin{abstract}
For strongly correlated electrons the  relation between  total number  of charge 
carriers $n_e$ and the chemical potential $\mu$ reveals   for large Coulomb energy
the apparently paradoxical pinning of $\mu$ within the Mott gap, as observed in high-$T_c$ cuprates. By unravelling consequences of the   non-trivial topology of the charge  gauge U(1) group
and the associated ground state degeneracy
we found a close kinship between the pinning of $\mu$ and the zero-temperature
divergence of the charge compressibility
$\kappa\sim\partial n_e/\partial\mu$, which marks a  novel quantum criticality
governed by  topological charges rather than Landau principle of  the symmetry breaking.
\end{abstract}
\pacs{74.20.-z, 74.20.Mn, 74.72.-h}

\author{T. K. Kope\'{c}}
\affiliation{
Institute for Low Temperature and Structure Research, Polish Academy of Sciences,
POB 1410, 50-950 Wroclaw 2, Poland}
\maketitle
The location of the chemical potential $\mu$ is a fundamental issue  for theories of
doped  high-$T_c$ cuprates.
However, the understanding of the observed doping dependence
of $\mu$  in  cuprates remains a puzzle.
In the case of LSCO, the angle-resolved photoemission
spectroscopy  studies \cite{chem1,chem2} have shown that in underdoped
samples, the chemical potential  is pinned 
above the top of the lower Hubbard band.
The photoemission
measurements of core levels also shows that $\mu$
does not move with hole doping in the underdoped
region \cite{chem3}.
The observed shift in the chemical potential as a function of the electron density $n_e$ is related to the
charge compressibility $\kappa$ or the charge susceptibility $\chi_c$ via
$\kappa^{-1} = (1/n_e)(\partial\mu/ \partial n_e)$ and $\chi_c=\partial n_e/\partial\mu$, respectively.
In Fermi liquid theory  (FLT) the parameter $\kappa$ is proportional to the effective mass $m^\star$ of the quasiparticles,
thus the absence of a shift  ($\kappa\to \infty$), quite peculiar
from the viewpoint of the rigid-band model, would imply the divergence of the
effective mass. The  pinning behavior of
$\mu$   can be made plausible for a system
where  charge carriers  are segregated, e.g., in a stripe form \cite{emery}.
However, it is hard to imagine how  the property $m^\star\to \infty $ 
might arise due to inhomogeneities, since the  mass enhancement  appears to result
from the physics of the uniform liquid \cite{sarma}. Anyway, the  behavior of $\mu$
is quite peculiar from the viewpoint of the FLT of the metallic states and signals
a dramatic reorganization of the electronic structure of cuprates  with  doping.

Given the fact that the electron density appears (besides the  temperature $T$) as a control
parameter, it follows from scaling and  thermodynamical considerations that
the  existence of quantum phase transitions \cite{sachdev}
and associated zero--temperature critical points might be a part of the physics of high-$T_c$ cuprates.
Indeed, in cuprates there is clear evidence for the existence of a special
doping point in the lightly-overdoped region  where superconductivity is
most robust. Such behaviour indicates this point could be a  quantum critical point
(QCP) while  the associated critical fluctuations might be responsible for the 
unconventional normal state behaviour \cite{qcp}. However, 
it is unclear whether this QCP is ``truly critical" in the sense
that it is characterized by universality and hyperscaling.
The resemblance to a conventional QCP is hampered by the lack of any clear signature of
thermodynamic critical behavior.
Experiments appear to exclude any broken  symmetry 
around this point  although  a sharp change in transport properties is 
observed \cite{transport}.

Understanding these novel phenomena in cuprates are challenging tasks.
In the celebrated Hubbard model, which epitomizes the  character of strongly correlated
electrons the degeneracy of
energy states renders the usual perturbation theory inapplicable.
However, there is an intimate relation between 
the  ground state degeneracy and the topology of the system-
much of the quantum information 
is encoded in {\it phase} relations of the many-body wave function \cite{geophase}.
The non-integrable Berry phase factors \cite{berry} of the topological origin  that
 the eigenvectors of the Hubbard Hamiltonian 
possesses can produce the interference  akin to the Aharonov-Bohm
effect  \cite{ab} in an electromagnetic field giving rise to more
exotic elementary excitations \cite{wilczek}.

In this work, we explore Mott transitions from  the  non-magnetic insulator to a superconductor induced by doping and show that the process is governed by the  topological structure of the electromagnetic compact gauge U(1) group. As a result  collective 
instanton excitations of the phase field (dual to the charge) arise with a great degree of stability, governed by gauge flux changes by an integer multiples of $2\pi$,
which  labels the ground state  degeneracy. The associated abrupt transition
between differnt ``vacua" allows us to make link between the
unusual behavior of the chemical potential and a novel type of
quantum criticality that goes beyond the paradigm of the  symmetry breaking.

We consider an effective one--band electronic
Hamiltonian on a tetragonal lattice that emphasises strong anisotropy and
the presence of a layered CuO$_2$ stacking sequence in cuprates:
${\cal H}= {\cal H}_{t-J}+{\cal H}_U+{\cal H}_\perp$, where
\begin{eqnarray}
&&{\cal H}_{t-J}=\sum_{\alpha\ell}
\left[ -\sum_{\langle {\bf r}{\bf r}'\rangle}
 tc^{\dagger }_{{\alpha}\ell}({\bf r})
c_{\alpha \ell}({\bf r}')
\right.
\nonumber\\
&&+
\left.
\sum_{{\langle \langle{\bf r}{\bf r}'\rangle\rangle}}
 t'c^{\dagger }_{{\alpha}\ell}({\bf r})
c_{\alpha \ell}({\bf r}') -\mu\sum_{ {\bf r}}
 c^{\dagger }_{{\alpha}\ell}({\bf r})
c_{\alpha \ell}({\bf r})  \right]
\nonumber\\
&&+\sum_\ell\sum_{{\langle {\bf r}{\bf r}'\rangle}}
{J}\left[{\bf S}_\ell{({\bf r})}
\cdot{\bf S}_\ell{({\bf r}')}
-\frac{1}{4}{n}_\ell{({\bf r})}{n}_\ell{({\bf r}')}\right].
\label{tj}
\end{eqnarray}
Here $\langle {\bf r},{\bf r}'\rangle$ and 
$\langle\langle {\bf r},{\bf r}'\rangle\rangle$
denotes  summation
over the nearest-neighbour and next--nearest--neighbour
sites labelled by $1\le {\bf r}\le N$ within the CuO  plane, respectively, with $t,t'$ being the {\it bare} 
hopping integrals  $t'>0$, while $1\le\ell\le N_\perp$ labels copper-oxide layers.
The operator $c_{\alpha\ell}^\dagger({\bf r})(c_{\alpha\ell}({\bf r}))$
creates (annihilates) an electron of spin $\alpha$ at the lattice site $({\bf r},\ell)$,
$S^a_{\ell}({\bf r})=\sum_{\alpha\beta}c^\dagger_{\alpha\ell}({\bf r})
\sigma_{\alpha\beta}^a c_{\beta\ell}({\bf r})$ ($a=x,y,z$) stands for spin
and ${n}_\ell({{\bf r}})= n_{\uparrow\ell} ({\bf r})+n_{\downarrow\ell}({\bf r})$ number operators,
respectively, where  $ {n}_{\alpha\ell}({{\bf r}})= c^\dagger_{\alpha\ell}({{\bf r}})
c_{\alpha\ell}({\bf r})$, $\mu$ is the chemical potential
and $J$ the antifferomagnetic (AF) exchange.
The Hubbard term is ${\cal H}_U=\sum_{\ell\bf r}
Un_{\uparrow\ell} ({\bf r}) n_{\downarrow\ell}({\bf r})$ with the on--site repulsion Coulomb energy $U$,
while ${\cal H}_\perp=-
\sum_{{\bf r}{\bf r}'}
t_\perp({\bf r}{\bf r}')
 c^{\dagger }_{{\alpha}\ell}({\bf r})
c_{\alpha \ell+1}({\bf r}')$ facilitates the interlayer coupling,
where $t_\perp$ is the interlayer hopping
with the $c-$axis dispersion  $\epsilon_\perp({\bf k},k_z)=2t_\perp({\bf k})\cos(ck_z)$,
while $t_\perp({\bf k})=t_\perp\left[\cos(ak_x)-\cos(ak_y)\right]^2$
 as predicted on the basis of  band calculations \cite{ander}.
We write the partition function $Z=\int\left[{\cal D}\bar{c}  {\cal D}\bar{c}
\right]e^{-{\cal S}[\bar{c},c]}$ 
with the action 
\begin{eqnarray}
{\cal S}[\bar{c},c]=\int_0^\beta d\tau[\sum_{\alpha{\bf r}\ell}
 \bar{c}_{\alpha\ell}({\bf r}\tau)\partial_\tau{c}_{\alpha\ell}({\bf r}\tau)+{\cal H}(\tau)]
 \end{eqnarray}
using  coherent-state fermionic path integral
over Grassmann fields 
depending on the ``imaginary time" $0\le\tau\le \beta\equiv 1/k_BT$. 
Further, we write  the Hubbard term in a SU(2) invariant way as
${\cal H}_U(\tau)=U\sum_{{\bf r}\ell}\{({1}/{4}){n_\ell}^2({{\bf r}}\tau)
-\left[{\bf \Omega}_\ell({\bf r}\tau)\cdot{\bf S}_\ell({\bf r}\tau)\right]^2\}$ 
with charge-U(1)  and  spin-SU(2)/U(1)   sectors, where the unit vector
${\bf \Omega}_\ell({\bf r}\tau)$ sets
varying in space-time  spin quantization axis \cite{weng}. 
In the following we fix our attention on the
U(1) invariant {\it charge }sector \cite{kopec}.
We consider now the following resolution of unity using the Fadeev-Popov method \cite{popov},
$1\equiv\int[{\cal  D} Q]\delta\left[Q-n\right]=
\int\left[\frac{{\cal D}V }{2\pi}  \right]e^{i\sum_{\bf r}\int_0^\beta d\tau (Q-n)V}$,
where $n({\bf r}\tau)=\sum_\alpha \bar{c}_\alpha({\bf r}\tau) c_\alpha({\bf r}\tau)$
represents the particle number, $Q({\bf r}\tau)$ is the collective variable and
 $iV({\bf r}\tau)$ is fluctuating (in space and time) imaginary chemical potential 
conjugate to the local particle number $n_\ell({\bf r}\tau)$ .
The field  $V_\ell({\bf r}\tau)$ can be written as a sum of
a static $V_{0\ell}({\bf r})$ and periodic function
$V({\bf r}\tau)=V_0({\bf r})+\tilde{V}({\bf r}\tau)$ using  Fourier series
$\tilde{V}({\bf r}\tau)=({1}/{\beta})\sum_{n=1}^\infty
[\tilde{V}({\bf r}\omega_n)e^{i\omega_n\tau}+c.c.]$
with $\omega_n=2\pi n/\beta$ ($n=0,\pm1,\pm2$)
being the (Bose) Matsubara frequencies.
Now, we introduce the {\it phase } (or ``flux") field ${\phi}_\ell({\bf r}\tau)$ via the Faraday--type relation
$\dot{\phi}_\ell({\bf r}\tau)\equiv\frac{\partial\phi_\ell({\bf r}\tau)}
{\partial\tau}=\tilde{V}_\ell({\bf r}\tau)$
to remove the imaginary term
 $i\int_0^\beta d\tau\dot{\phi}_\ell({\bf r}\tau) n_\ell({\bf r}\tau)
\equiv i\int_0^\beta d\tau\tilde{V}_\ell({\bf r}\tau)n_\ell({\bf r}\tau)$ for all the Fourier modes
of the $V_\ell({\bf r}\tau)$ field, except for $\omega_n=0$
by performing the  gauge transformation to the {\it new} fermionic variables $f_{\alpha\ell}({\bf r}\tau)$, where
$c_{\alpha\ell}({\bf r}\tau)=e^{i{\phi}_\ell({\bf r}\tau)}f_{\alpha\ell}({\bf r}\tau)$,which indicates that
the electron acquire a phase shift similar to that
in the 	electric  AB effect \cite{ab}.
 The electromagnetic  U(1)
group governing the phase field is {\it compact}, {\it i.e.}
$\phi_\ell({\bf r}\tau)$  has the topology of a circle ($S_1$), so that
topological effects can arise due to  non-homotopic 
mappings of the configuration space onto the gauge group $S_1\to$ U(1). To explicate the composite  nature
of the physical electron field we write explicitly:
\begin{eqnarray}
c_{\alpha\ell}({\bf r}\tau)=
\exp\left[ i\int_0^\tau d\tau' \tilde{V}_{D\ell}({\bf r}\tau')\right]e^{i\gamma_{\ell B}({\bf r}\tau)}
f_{\alpha\ell}({\bf r}\tau).
\label{compo}
\end{eqnarray}
The first term in the exponential in Eq.(\ref{compo}) is the usual dynamical phase factor where 
$\dot{\theta}_\ell({\bf r}\tau)=\tilde{V}_{D\ell}({\bf r}\tau)$ and
${\theta}_\ell({\bf r}\beta)=\theta_\ell({\bf r}0)$ .
The second one is the {\it non-integrable} Berry phase factor \cite{berry}:
$\gamma_{\ell B}({\bf r}\tau)=2\pi\tau m_\ell({\bf r})/\beta$,
where $m_\ell({\bf r})$ marks the integer U(1) {\it winding }  number. 
Working backwards we can now formulate the the path integral for the partition function. For this we concentrate on 
closed paths (or world lines)in the imaginary time $\tau$ that start at position ${\bf r}\ell$
 at imaginary time $\tau=0$ and end at
the same position  at $\tau=\beta$,
 which fall into distinct, disconnected (homotopy) classes
labelled by the  winding number \cite{schulman}. Homotopically distinct paths can be summed
according to  various possibilities for inequivalent
quantizations (superselection sectors) according to the formula:
\begin{widetext}
\begin{eqnarray}
Z&=&\sum_{ \{m_\ell({\bf r})\}}\int_0^{2\pi}\prod_{{\bf r}\ell}d\phi_{0\ell}({\bf r})
\int_{\phi_\ell({\bf r}0)=\phi_{0\ell}({\bf r})}^{\phi_\ell({\bf r}\beta)
=\phi_{\ell 0}({\bf r})+2\pi m_\ell({\bf r})}
\prod_{{\bf r}\ell\tau}{d}\theta_\ell({\bf r}\tau) \int\left[{\cal D}\bar{f}  {\cal D}{f}
\right]e^{-{\cal S}[{\theta},m,\bar{f},f]},
\nonumber\\
{\cal S}[{\theta},m,\bar{f},f]&=&\sum_{ \ell}
\int_0^\beta d\tau\left\{
\frac{1}{U}\sum_{\bf r}\left[ 
\frac{\partial\theta_\ell({\bf r}\tau)}{\partial\tau}+\frac{2\pi}{\beta} m_\ell({\bf r})\right]^2
+\frac{2\mu}{U}\sum_{ {\bf r}}
\frac{1}{i}\left[ 
\frac{\partial\theta_\ell({\bf r}\tau)}{\partial\tau}+\frac{2\pi}{\beta} m_\ell({\bf r})\right]
\right.
\nonumber\\
&+&\left.H\left[   \bar{f}_{\alpha\ell}({\bf r}\tau),{f}_{\alpha\ell}
({\bf r}\tau),\theta_\ell({\bf r}\tau), m_\ell({\bf r})\right]\right\}.
\label{explicit}
\end{eqnarray}
\end{widetext}
The   first order in time derivative term in Eq.(\ref{explicit}) is just the topological
action which does not affect the equation of motions
but influences  statistics \cite{wilczek}.
The  gauge transformation in Eq.(\ref{compo}) introduces phase factor 
into the hopping elements of the hamiltonian which frustrate the motion of the
fermionic subsystem. However,  when charge  fluctuations become {\it phase coherent}
is  signalled by $\langle e^{i\phi_\ell({\bf r}\tau)}\rangle\neq 0$
the frustration of the kinetic energy is released.  
To proceed, we trace over
the fermionic degrees of freedom in Eq.(\ref{explicit}) and
 introduce  the unimodular complex scalar
 $z_\ell({\bf r}\tau)=e^{i\phi_\ell({\bf r}\tau)}$ via another Fadeev-Popov resolution of unity
\begin{eqnarray}
1&\equiv&\int\left[{\cal D}^2 {z}\right]\prod_{{\bf r}\ell}
\delta\left(|{z_\ell({\bf r}\tau)}|^2-1\right)\times
\nonumber\\
&\times & \Re\left[ z_\ell({\bf r}\tau)-e^{i\phi_\ell({\bf r}\tau)}
 \right]\Im\left[ z_\ell({\bf r}\tau)-e^{i\phi_\ell({\bf r}\tau)} \right],
\end{eqnarray}
where the unimodularity constraint can be imposed
with a real Lagrange multiplier  $\lambda$. The partition function then becomes
$Z=\int\left[{\cal D}^2 {z}\right]\prod_{{\bf r}\ell}
\delta\left(|{z_\ell({\bf r}\tau)}|^2-1\right)e^{-{\cal S}[z,z^\star]}$, where
${\cal S}[z,z^\star]=
\frac{1}{\beta N N_\perp}\sum_{{\bf q}\omega_n}
{z}_{\bf q}^\star(\omega_n)\Gamma^{-1}_{\bf q}(\omega_n){z}_{\bf q}(\omega_n)$.
The phase-coherence boundary  is
$1 = \left.\frac{1}{\beta N N_\perp}\sum_{{\bf q}\omega_n}
{ \Gamma}_{\bf q}(\omega_n)\right|_{\lambda=\lambda_c}$
Here, ${\bf q}\equiv({\bf k},k_z)$  and 
at criticality, the condition  $\Gamma^{-1}_{{\bf q}=\bf 0}(\omega_n=0)|_{\lambda=\lambda_c}=0$ fixes the Lagrange
parameter while
\begin{eqnarray}
&&\Gamma^{-1}_{\bf q}(\omega_n)=\lambda-\Sigma({\bf q},\omega_n)
+\gamma^{-1}(\omega_n)
\nonumber\\
&&\Sigma({\bf q},\omega_n)=
{\cal J}'_\|(\Delta)\cos(ak_x)\cos(ak_y)
\nonumber\\
&&+\sum_{\bf q'}\frac{\xi_{\bf k'}\Gamma_{{\bf q}+{\bf q'}}(\omega_n)}{NN_\perp}
\left[ \bar{\cal J}_\|(\Delta)+{\cal J}_\perp(\Delta)\cos(ck_z) 	\right],
\end{eqnarray}
where $\xi_{\bf k}=\cos(ak_x)+\cos(ak_y)$
Furthermore, $\gamma_{0}(\omega_n)$ is the Fourier transform of the
bare phase correlator 
$\langle e^{-i[\phi_\ell({\bf r}\tau)-\phi_{\ell'}({\bf r}'\tau')]}\rangle_0$
originating from the kinetic and  topological part of the action in Eq.(\ref{explicit})
\begin{eqnarray}
\gamma(\omega_n)=\frac{1}{Z_0}\sum_{m=-\infty}^{+\infty}
\frac{\frac{8}{U}\exp\left[-\frac{\beta U}{4}\left(m-\frac{2\mu}{U}\right)^2  \right]}
{ 1-4	\left[\left(m-\frac{2\mu}{U} \right)
-\frac{2i\omega_n }{U} \right]^2},
\end{eqnarray}
where $Z_0=\exp(-\frac{\beta \mu^2}{U})\theta_3(\frac{\beta\mu}{2\pi i}, 
e^{-{\beta U}/{4} })$ and $\theta_3(v,q)=\sum_{m=-\infty}^{+\infty}q^{m^2}e^{2\pi m iv}$ is the Jacobi theta function 
\cite{EllipticFunction}
which is $\beta$-periodic in the
"imaginary-time" $\tau$ as well as in the variable $2\mu/U$ with the period of unity.
The microscopic {\it phase stiffnesses} to the lowest order in the hopping amplitudes
 are given by
\begin{eqnarray}
{\cal J}_\|(\Delta)&=&\frac{1}{2}\frac{t^2}{\beta N^2}\sum_{\nu_n}\left[\sum_{{\bf k}}
\frac{|\Delta({\bf k})|}{\nu^2_n
+\mu^2+|\Delta({\bf k})|^2}\right]^2,
\nonumber\\
{\cal J}'_\|(\Delta)&=&-\frac{t'\bar{\mu}}{\beta N}\sum_{{\bf k},\nu_n}
\frac{\cos(ak_x)\cos(ak_y) }{\nu^2_n
+\bar{\mu}^2+|\Delta({\bf k})| },
\nonumber\\
{\cal J}_\perp(\Delta)&=&\frac{1}{\beta N^2}\sum_{{\bf k'}{\bf k}}
\sum_{\nu_n}\frac{t^2_\perp({\bf k'})|\Delta({\bf k})|^2}{\left[\nu^2_n
+\mu^2+|\Delta({\bf k})|^2\right]^2},
\label{stiff}
\end{eqnarray}
where $\bar{\mu}=\mu-n_f{U}/{2}$, $n_f=\langle \bar{f}_{\alpha}({\bf r}\tau)
f_{\alpha\ell}({\bf r}\tau)\rangle$ is the occupation number for $f-$fermions
and $\nu_n=\pi(2n+1)/\beta$ ($n=0,\pm1,\pm2$) stand for the Fermi Matsubara frequency.
The stiffnesses in Eq.(\ref{stiff})
rest on the ``$d$-wave" pair amplitude  $\Delta({\bf k})=|\Delta|\eta_{\bf k}$ where $\eta_{\bf k} =\cos(ak_x)-\cos(ak_y)$
 due to the  in-plane  momentum space   pairing of the $f$-fermions. 
 A Gorkov-type decoupling of the  AF exchange term in Eq.(\ref{tj})  using
the valence bond operator \cite{bond} readily gives for the gap parameter:	
$1=\frac{J}{N}\sum_{\bf k}\frac{\eta^2_{\bf k}}{2{\cal E}_{\bf k}}
\tanh\left[
\frac{1}{2}\beta{\cal E}_{\bf k} \right]$
with the quasiparticle spectrum, ${\cal E}^2_{\bf k}=[\epsilon_\|^\star({\bf k})-\bar{\mu}]^2
+|\Delta({\bf k})|^2$. Here, $\epsilon_\|^\star({\bf k})$ is effective  in-plane band dispersion
narrowed due to the  frustrated
motion of the carriers in the fluctuating ``bath" of U(1) gauge potentials, so that
the actual tight-binding parameters are ``dressed" ones
$t^\star_X=t_X \langle e^{-i[\phi_\ell({\bf r}\tau)-\phi_\ell({\bf r}'\tau)]}\rangle$, where 
$t_X=t,t',t_\perp$ are the bare band parameters. 
\begin{figure}
\begin{center}
\includegraphics[width=7.5cm]{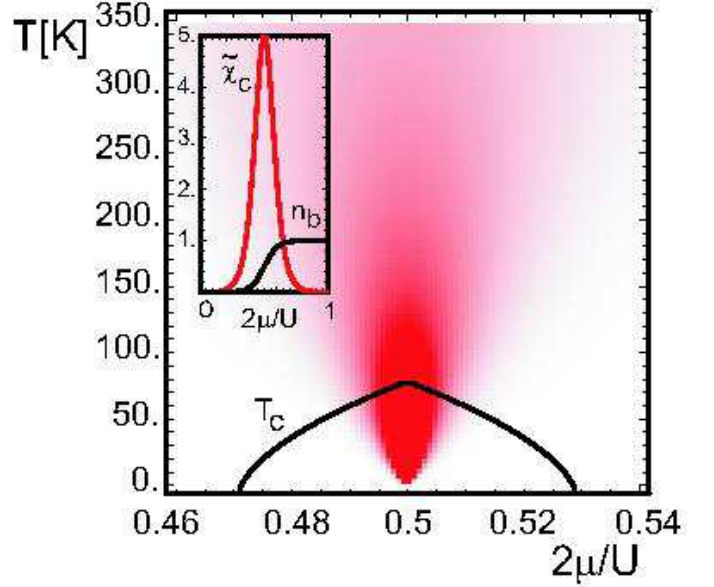}
\end{center}
 \caption{(Color online) The critical temperature $T_c$ as a function of
the chemical potential $\mu$ together with the density plot of the  charge susceptibility 
$\tilde {\chi}_c\equiv U\chi_c/2$ for $t^\star=0.5$eV, $t'^\star/t^\star=0.3$, $t^\star_\perp=0.01$eV,
 $J=0.15$eV and $U=4$eV.
Inset: $\tilde {\chi}_c$ and the occupation number $n_b$ for $T=0.1U$.
   }\label{fig1}
\end{figure} 
Now, by differentiating the free energy  before after implementing of the
gauge transformation in Eq.(\ref{compo}) we obtain for the electron occupation number
$n_e\equiv\langle \bar{c}_{\alpha}({\bf r}\tau)c_{\alpha\ell}({\bf r}\tau)\rangle$ the result
$n_e=n_f+ n_b-{2\mu}/{U}$. Here, $n_b={2\mu}/{U}+({2}/{iU})\langle\dot{\phi}({\bf r}\tau)\rangle$
is the occupation number given by the mean value of the winding numbers 
\begin{eqnarray}
n_b(\mu)=n_b^0(\mu)+\frac{1}{\beta N N_\perp}\sum_{{\bf q}\omega_n}
{ \Gamma}_{\bf q}(\omega_n)\partial_\mu \gamma^{-1}(\omega_n)
\end{eqnarray}
with
\begin{eqnarray}
n_b^0(\mu)&=&\frac{2\mu}{U} -\frac{1}{\beta} \frac{\partial_\mu\theta_3\left[\frac{\beta\mu}{2\pi i}, 
e^{-{\beta U}/{4} }\right]}{\theta_3\left[\frac{\beta\mu}{2\pi i}, 
e^{-{\beta U}/{4} }\right]},
\end{eqnarray}
where we made use of the relation
\begin{eqnarray}
&&\frac{\partial_v\theta_3(iv,q)}{\theta_3(iv,q)}
=\sum_{m=1}^{\infty}\frac{(-1)^m4\pi iq^m}{1-q^{2m}}\sinh(2\pi m v).
 \end{eqnarray}
%
\begin{figure}
\begin{center}
\includegraphics[width=7.5cm]{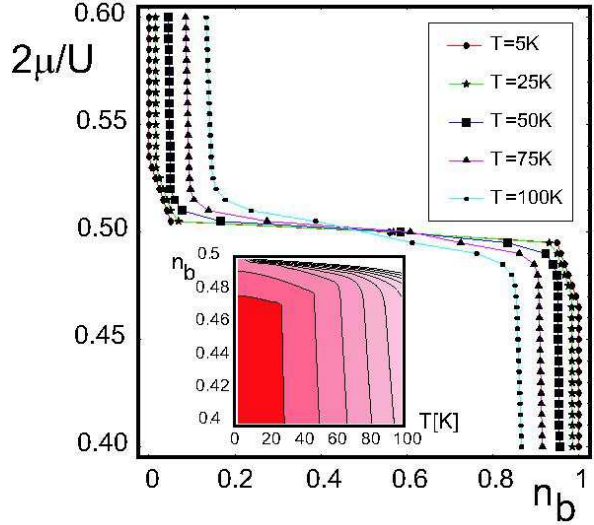}
\end{center}
 \caption{(Color online) The chemical potential $\mu$ as a function of the
occupation number $n_b$ for various temperatures $T$ as indicated in the plot and model's
parameters as in Fig.1. The value of $\mu$ stays within the charge gap
as  $n_b$ changes.  Inset: the density plot of $n_b$ as a function 
of $\mu$ and $T$.
   }\label{fig2}
\end{figure} 
%
In the limit of strong (weak) correlations $n_e$ interpolates between topological $n_b$ (fermionic $n_f$) occupation numbers.
In the large--$U$ limit 
$\mu\to n_fU/2$, so that $n_e\to n_b$  and the system behaves as governed entirely by
U(1) topological charges which play the role of ``quasiparticles".
It is straightforward now to calculate the charge compressibility $\kappa$. The result is given
in Fig.1 along with the outcome for the superconducting phase boundary.
We see the evolution of $\kappa$ with decreasing $n_e$, (i.e. hole doping) from the Mott insulator \cite{mott}
with $\kappa=0$ (at $2\mu/U=1$) to a point of degeneracy  on the brink of the
particle occupation change at $2\mu/U=.5$ where $\kappa=\infty$ at $T=0$.
This is also the point on the phase diagram   from which the superconducting lobe emanates.
It is clear that, the nature of the divergence of $\kappa$ here has nothing to do
with singular fluctuations due to spontaneous symmetry breaking as in the ``conventional" phase transition.
Rather, this divergent response appears as a kind of topological protection
built in the system against
the small changes of $\mu$.
Further, $\kappa\to \infty$ implies that the and $\partial\mu/\partial n_e$
becomes vanishingly small  at $T=0$ which results in the  chemical potential pinning,
see Fig.2, where  the temperature dependence of $\mu$ is also shown in the inset.
In  cuprates  $\mu(T)$  can b accessed via the measurement
of the work function \cite{marel}: the available data for YBCO clearly indicates the temperature
behavior of $\mu$ characteristic for  a bosonic system.

To conclude, new type of quantum numbers must be invoked to explain topologically induced QCP in cuprates
and the associated pinning of the chemical potential.
Topological effects arise as stable, non-perturbative, collective excitations of the phase field
(dual to the charge), which  carry  novel topological characteristics.
These are the winding numbers of U(1) group:
$ m_\ell({\bf r})\equiv\frac{1}{2\pi}\int_0^\beta d\tau\dot{\phi}_\ell({\bf r}\tau)$ that
 become topological conserved quantities.
It is exactly the appearance of these topological charges
that render the system ``protected" against small changes of the hamiltonian's
parameters.
This novel conservation does not arise  just out of a
symmetry of the theory (as ``conventional" conservation laws based on Noether's theorem) 
but it is a consequence of the connectedness, i.e. topology 
of the phase  space,  related to the topological properties of the associated symmetry group.

This work was supported by the Polish Science Committee (KBN)
under the grant No. 2P03B 009 25 in years 2003-2005.


\begin{references}
\bibitem{chem1}
%
 A. Ino, C. Kim, M. Nakamura, T. Yoshida, T. Mizokawa, Z.-X.
Shen, A. Fujimori, T. Kakeshita, H. Eisaki, and S. Uchida, Phys.
Rev. B {\bf 62}, 4137 (2000).
%
\bibitem{chem2}
A. Ino, C. Kim, M. Nakamura, T. Yoshida, T. Mizokawa, A. Fujimori,
Z.-X. Shen, T. Kakeshita, H. Eisaki, and S. Uchida, Phys.
Rev. B {\bf 65}, 094504 (2002).
%
\bibitem{chem3}
A. Ino, T. Mizokawa, A. Fujimori, K. Tamasaku, H. Eisaki, S.
Uchida, T. Kimura, T. Sasagawa, and K. Kishio, Phys. Rev. Lett.
{\bf 79}, 2101 (1997).
%
\bibitem{emery}
V. J. Emery and S. A. Kivelson, J. Phys. Chem. Solids {\bf 53},
1499 (1992).

\bibitem{sarma}
Y. Zhang and S. Das Sarma, Phys. Rev. B{\bf 70}, 035104 (2004).
%
\bibitem{sachdev}
S. Sachdev, {\it Quantum Phase Transitions} (Cambridge
Univ. Press, Cambridge, 1999); M. Vojta, Rep. Prog.
Phys. {\bf 66}, 2069 (2003).
%
\bibitem{qcp}
C.M.Varma, Phys. Rev. B{\bf  55}, 14554 (1997);
S.H. Naqib, J.R. Cooper, J.L. Tallon and C. Panagopoulos, cond-mat/0301375.
%
\bibitem{transport}
T. Ito, K. Takenaka, and S. Uchida, Phys. Rev. Lett. {\bf 70}, 3995 (1993).
%
\bibitem{berry}
M. V. Berry, Proc. R. Soc. London, Ser. {\bf A} 392, 451 (1984).
%
\bibitem{geophase}
A. Shapere and F. Wilczek, {\it Geometric phases in physics}, (World Scientific, Singapore, 1989)
%
\bibitem{ab}
Y. Aharonov and D. Bohm, Phys. Rev. 115, 485 (1959).
%
\bibitem{wilczek}
F. Wilczek, {\it Fractional Statistics and Anyon Superconductivity},
(World Scientific, Singapore, 1990).
%
\bibitem{ander}
O. K. Andersen, A. I. Liechtenstein, O. Jepsen, and F. Paulsen,  J. Phys. Chem. Solids {\bf 56}, 1573 (1995).
%
\bibitem{kopec}
T. K. Kope\'c, Phys. Rev. B{\bf 70}, 054518 (2004).
%
\bibitem{popov}
L. D. Faddeev and V. N. Popov, Phys. Lett. {\bf 25}, 29 (1967).
%
\bibitem{weng}
Z. Y. Weng and C. S. Ting and T. K. Lee, Phys. Rev. B {\bf 43}, R3790 (1991).
%
\bibitem{schulman}
L. S. Schulman, {\it Techniques and Applications of Path
Integration} (Wiley, New York 1981).
%
\bibitem{EllipticFunction}  M. Abramovitz and I. Stegun, {\it Handbook of
Mathematical Functions }(Dover, New York, 1970).

\bibitem{bond}
G. Baskaran, Z. Zou, and P. W. Anderson, Solid State Commun.
{\bf 63}, 973 (1987).
%
\bibitem{mott}
N. F. Mott, {\it Metal-Insulator Transitions} (Taylor \& Francis, London, 1990).
%
\bibitem{marel}
G. Rietveld, N. Y. Chen, and D. van der Marel,Phys. Rev. Lett. {\bf 69}, 2578 (1992)
%

\end{references}
 \end{document}